# High-Performance Concurrency Control Mechanisms for Main-Memory Databases

Per-Åke Larson[1], Spyros Blanas[2], Cristian Diaconu[1],
Craig Freedman[1], Jignesh M. Patel[2], Mike Zwilling[1]
Microsoft[1], University of Wisconsin – Madison[2]
{palarson, cdiaconu, craigfr,mikezw}@microsoft.com, {sblanas, jignesh}@cs.wisc.edu

## ABSTRACT

A database system optimized for in-memory storage can support much higher transaction rates than current systems. However, standard concurrency control methods used today do not scale to the high transaction rates achievable by such systems. In this paper we introduce two efficient concurrency control methods specifically designed for main-memory databases. Both use multiversioning to isolate read-only transactions from updates but differ in how atomicity is ensured: one is optimistic and one is pessimistic. To avoid expensive context switching, transactions never block during normal processing but they may have to wait before commit to ensure correct serialization ordering. We also implemented a main-memory optimized version of single-version locking. Experimental results show that while single-version locking works well when transactions are short and contention is low performance degrades under more demanding conditions. The multiversion schemes have higher overhead but are much less sensitive to hotspots and the presence of long-running transactions.

## 1. INTRODUCTION

Current database management systems were designed assuming that data would reside on disk. However, memory prices continue to decline; over the last 30 years they have been dropping by a factor of 10 every 5 years. The latest Oracle Exadata X2-8 system ships with 2TB of main memory and it is likely that we will see commodity servers with multiple terabytes of main memory within a few years. On such systems the majority of OLTP databases will fit entirely in memory, and even the largest OLTP databases will keep the active working set in memory, leaving only cold, infrequently accessed data on external storage.

A DBMS optimized for in-memory storage and running on a many-core processor can support very high transaction rates. Efficiently ensuring isolation between concurrently executing transactions becomes challenging in such an environment. Current DBMSs typically rely on locking but in a traditional implementation with a separate lock manager the lock manager becomes a bottleneck at high transaction rates as shown in experiments by Johnson et al [15]. Long read-only transactions are also problematic as readers may block writers.

This paper investigates high-performance concurrency control mechanisms for OLTP workloads in main-memory databases. We found that traditional single-version locking is "fragile". It works well when all transactions are short and there are no hotspots but performance degrades rapidly under high contention or when the workload includes even a single long transaction.

Decades of research has shown that multiversion concurrency control (MVCC) methods are more robust and perform well for a broad range of workloads. This led us to investigate how to construct MVCC mechanisms optimized for main memory settings. We designed two MVCC mechanisms: the first is optimistic and relies on validation, while the second one is pessimistic and relies on locking. The two schemes are mutually compatible in the sense that optimistic and pessimistic transactions can be mixed and access the same database concurrently. We systematically explored and evaluated these methods, providing an extensive experimental evaluation of the pros and cons of each approach. The experiments confirmed that MVCC methods are indeed more robust than single-version locking.

This paper makes three contributions. First, we propose an optimistic MVCC method designed specifically for memory resident data. Second, we redesign two locking-based concurrency control methods, one single-version and one multiversion, to fully exploit a main-memory setting. Third, we evaluate the effectiveness of these three different concurrency control methods for different workloads. The insights from this study are directly applicable to high-performance main memory databases: single-version locking performs well only when transactions are short and contention is low; higher contention or workloads including some long transactions favor the multiversion methods; and the optimistic method performs better than the pessimistic method.

The rest of the paper is organized as follows. Section 2 covers preliminaries of multiversioning and describes how version visibility and updatability are determined based on version timestamps. The optimistic scheme and the pessimistic scheme are described in Section 3 and Section 4, respectively. Section 5 reports performance results. Related work is discussed in Section 6, and Section 7 offers concluding remarks. Proofs of correctness are provided in an online addendum to this paper and at [27].

## 2. MV STORAGE ENGINE

A transaction is by definition serializable if its reads and writes logically occur as of the same time. The simplest and most widely used MVCC method is snapshot isolation (SI). SI does not guarantee serializability because reads and writes logically occur at different times, reads at the beginning of the transaction and writes at the end. However, a transaction is serializable if we can guarantee that it would see exactly the same data if all its reads were repeated at the end of the transaction.

To ensure that a transaction T is serializable we must guarantee that the following two properties hold:

1. **Read stability.** If T reads some version V1 of a record during its processing, we must guarantee that V1 is still the version visible to T as of the end of the transaction, that is, V1 has not been replaced by another *committed* version V2. This can be implemented either by read locking V1 to prevent updates or by validating that V1 has not been updated before commit. This ensures that nothing has disappeared from the view.
2. **Phantom avoidance.** We must also guarantee that the transaction's scans would not return additional new versions. This can be implemented in two ways: by locking the scanned part of an index/table or by rescanning to check for new versions before commit. This ensures that nothing has been added to the view.

Lower isolation levels are easier to support.
- For repeatable read, we only need to guarantee read stability.
- For read committed, no locking or validation is required; always read the latest committed version.
- For snapshot isolation, no locking or validation is required; always read as of the beginning of the transaction.

We have implemented a prototype main-memory storage engine. We begin with a high-level overview of how data is stored, how reads and updates are handled, and how it is determined what versions are visible to a reader, and that a version can be updated.

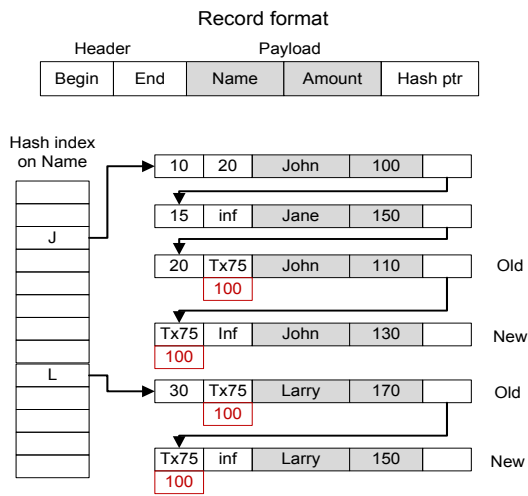

**Figure 1:** Example account table with one hash index. Transaction 75 has transferred $20 from Larry's account to John's account but has not yet committed.

## 2.1 Storage and Indexing

Our prototype currently supports only hash indexes which are implemented using lock-free hash tables. A table can have many indexes, and records are always accessed via an index lookup - there is no direct access to a record without going through an index. (To scan a table, one simply scans all buckets of any index on the table.) The techniques presented here are general and can be applied to ordered indexes implemented by trees or skip lists.

Figure 1 shows a simple bank account table containing six versions. Ignore the numbers (100) in red for now. The table has two (user) columns: Name and Amount. Each version has a valid time defined by timestamps stored in the Begin and End fields. A table can have several indexes, each one with a separate (hash) pointer field. The example table has one index on the Name column. For simplicity we assume that the hash function just picks the first letter of the name. Versions that hash to the same bucket are linked together using the Hash ptr field.

Hash bucket J contains four records: three versions for John and one version for Jane. The order in which the records are linked together is immaterial. Jane's single version (Jane, 150) has a valid time from 15 to infinity meaning that it was created by a transaction that committed at time 15 and it is still valid. John's oldest version (John, 100) was valid from time 10 to time 20 when it was updated. The update created a new version (John, 110) that initially had a valid time of 20 to infinity. We will discuss John's last version (John, 130) in a moment.

## 2.2 Reads

Every read specifies a logical (as-of) read time and only versions whose valid time overlaps the read time are visible to the read; all other versions are ignored. Different versions of a record have non-overlapping valid times so at most one version of a record is visible to a read. A lookup for John, for example, would be handled by a scan of bucket J that checks every version in the bucket but returns only the one whose valid time overlaps the read time.

## 2.3 Updates

Bucket L contains two records which both belong to Larry. Transaction 75 is in the process of transferring $20 from Larry's account to John's account. It has created the new versions for Larry (Larry, 150) and for John (John, 130) and inserted them into the appropriate buckets in the index.

Note that transaction 75 has stored its transaction ID in the Begin and End fields of the new and old versions, respectively. (One bit in the field indicates the field's current content.) A transaction ID stored in the End field serves as a write lock and prevents other transactions from updating the same version and it identifies which transaction has updated it. A transaction Id stored in the Begin field informs readers that the version may not yet be committed and it identifies which transaction owns the version.

Now suppose transaction 75 commits with end timestamp 100. It then returns to the old and new versions and sets the Begin and End fields, respectively, to 100. The final values are shown in red below the old and new versions. The old version (John, 110) now has the valid time 20 to 100 and the new version (John, 130) has a valid time from 100 to infinity.

Every update creates a new version so we need to discard old versions that are no longer needed to avoid filling up memory. A version can be discarded when it is no longer visible to any transaction. In our prototype, once a version has been identified as garbage, collection is handled cooperatively by all threads. Although garbage collection is efficient and fully parallelizable, keeping track of the old versions does consume some processor resources. Details of our garbage collection algorithm are beyond the scope of this paper.

## 2.4 Transaction Phases

A transaction can be in one of four states: Active, Preparing, Committed, or Aborted. Fig. 2 shows the possible transitions between these states.

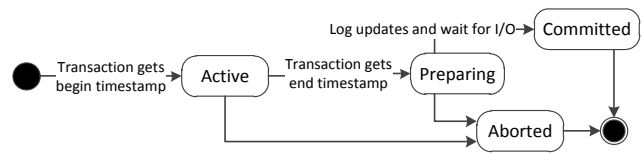

**Figure 2: State transitions for each transaction**

A transaction goes through three different phases. We outline the processing in each phase only briefly here; it is fleshed out in more detail in connection with each concurrency control method (Sections 3 and 4).

1. The transaction is created; it acquires a begin timestamp and sets its state to Active.
2. **Normal processing phase.** The transaction does all its normal processing during this phase. A transaction never blocks during this phase. For update operations, the transaction copies its transaction ID into the Begin field of the new versions and into the End field of the old or deleted versions. If it aborts, it changes its state to Aborted and skips directly to step 4. When the transaction has completed its normal processing and requests to commit, it acquires an end timestamp and switches to the Preparing state.
3. **Preparation phase.** During this phase the transaction determines whether it can commit or is forced to abort. If it has to abort, it switches its state to Aborted and continues to the next phase. If it is ready to commit, it writes all its new versions and information about deleted versions to a redo log record and waits for the log record to reach non-volatile storage. The transaction then switches its state to Committed.
4. **Postprocessing phase.** If the transaction has committed, it proceeds to replace its transaction ID with its end timestamp from the Begin field of the new versions and from the End field of the old or deleted versions. If the transaction has aborted, it marks all its new versions as garbage and sets their Begin and End timestamps to infinity to make them invisible.
5. The transaction is terminated. The old versions are assigned to the garbage collector, which is responsible for discarding them when they are no longer needed.

Timestamps are drawn from a global, monotonically increasing counter. A transaction's gets a unique timestamp by atomically reading and incrementing the counter.

## 2.5 Version Visibility

Under multiversioning a read must specify a logical read time. Only versions whose valid time overlaps the logical read time are visible to the read. The read time can be any value between the transaction's begin time and the current time. Which read time is chosen depends on the concurrency control method used and the transaction's isolation level; more about this in Sections 3 and 4.

While determining the visibility of a version is straightforward in principle, it is more complicated in practice as we do not want a transaction to block (wait) during normal processing. Recall that a version's Begin or End fields can temporarily store a transaction ID, if the version is being updated. If a reader encounters such a version, determining visibility without blocking requires checking another transaction's state and end timestamp and potentially even restricting the serialization order of transactions.

We now examine each case in turn, beginning from the easiest and most common case where both fields contain a timestamp, and then discussing the cases where the Begin or End fields contain transaction IDs.

*Begin and End fields contain timestamps*

Let RT denote the logical read time being used by a transaction T. To determine whether a version V is visible to T, we check V's Begin and End fields. If both fields contain timestamps, we simply check whether RT falls between the two timestamps. If it does, V is visible to T, otherwise not.

*Begin field contains a transaction ID*

Suppose transaction T reads version V and finds that V's Begin field contains the ID of a transaction TB. Version V may still be visible; it depends on transaction TB's state and TB's end timestamp. TB may, for example, have committed already but not yet finalized the Begin fields of its new versions. If so, V is a committed version with a well-defined begin time. Table 1 summarizes the various cases that may occur and the action to take depending on the state of the transaction TB.

Table 1: Case analysis of action to take when version V's Begin field contains the ID of transaction TB

| TB's state | TB's end timestamp | Action to take when transaction T checks visibility of version V. |
|---|---|---|
| Active | Not set | V is visible only if TB=T and V's end timestamp equals infinity. |
| Preparing | TS | V's begin timestamp will be TS but V is not yet committed. Use TS as V's begin time when testing visibility. If the test is true, allow T to *speculatively* read V. |
| Committed | TS | V's begin timestamp will be TS and V is committed. Use TS as V's begin time to test visibility. |
| Aborted | Irrelevant | Ignore V; it's a garbage version. |
| Terminated or not found | Irrelevant | Reread V's Begin field. TB has terminated so it must have finalized the timestamp. |

If TB is still in the Active state, the version is uncommitted and thus not visible to any other transactions than TB. If TB has updated a record multiple times, only the latest version is visible to it. V is the latest version if its End field contains infinity.

If transaction TB is in the Preparing state, it has acquired an end timestamp TS which will be V's begin timestamp if TB commits. A safe approach in this situation would be to have transaction T wait until transaction TB commits. However, we want to avoid all blocking during normal processing so instead we continue with the visibility test and, if the test returns true, allow T to *speculatively read* V. Transaction T acquires a commit dependency on TB, restricting the serialization order of the two transactions. That is, T is allowed to commit only if TB commits. Commit dependencies are discussed in more detail in Section 2.7.

The last three cases in Table 1 are straightforward.

*End field contains a transaction ID*

Once it has been determined that version V's valid time begins before transaction T's read time RT, we proceed to check V's End field. If it contains a timestamp, determining visibility is straightforward: V is visible to T if and only if RT is less than the timestamp. However, if the field contains the ID of transaction TE, we have to check the state and end timestamp of TE. Table 2 summarizes the various cases and the actions to take, assuming that we have already determined that V's begin timestamp is, or will be, less than RT.

The first case (TE is Active) was discussed earlier. If TE's state is Preparing, it has an end timestamp TS that will become the end timestamp of V if TE does commit. If TS is greater than the read

300

time RT, it is obvious that V will be visible if TE commits. If TE aborts, V will still be visible, because any transaction that updates V after TE has aborted will obtain an end timestamp greater than TS. If TS is less than RT, we have a more complicated situation: if TE commits, V will not be visible to T but if TE aborts, it will be visible. We could handle this by forcing T to wait until TE commits or aborts but we want to avoid all blocking during normal processing. Instead we allow T to *speculatively ignore* V and proceed with its processing. Transaction T acquires a commit dependency (see Section 2.7) on TE, that is, T is allowed to commit only if TE commits.

**Table 2: Case analysis of action to take when V's End field contains a transaction ID TE.**

| TE's state | TE's end timestamp | Action to take when transaction T checks visibility of a version V as of read time RT. |
|---|---|---|
| Active | Not set | V is visible only if V was created by transaction T and V's end timestamp equals infinity. |
| Preparing | TS | V's end timestamp will be TS provided that TE commits. If TS > RT, V is visible to T. If TS < RT, T *speculatively ignores* V. |
| Committed | TS | V's end timestamp will be TS and V is committed. Use TS as V's end timestamp when testing visibility |
| Aborted | Irrelevant | V is visible. |
| Terminated or not found | Irrelevant | Reread V's End field. TE has terminated so it must have finalized the timestamp. |

The case when TE's state is Committed is obvious but the Aborted case warrants some explanation. If TE has aborted, some other transaction TO may have sneaked in after T read V's End field, discovered that TE has aborted and updated V. However, TO must have updated V's end field after T read it and TO must have been in the Active state. This implies that TO's end timestamp was assigned after T read V and thus it must be later than T's logical read time, It follows that it doesn't matter if a transaction TO sneaked in; V is always visible to T if TE is in the Aborted state.

If TE has terminated or is not found, TE must have either committed or aborted and finalized V's end timestamp since we read the field. So we reread the End field and try again.

## 2.6 Updating a Version

Suppose transaction T wants to update a version V. V is updatable only if it is the latest version, that is, it has an end timestamp equal to infinity or its End field contains the ID of a transaction TE and TE's state is Aborted. If the state of transaction TE is Active or Preparing, V is the latest committed version but there is a later uncommitted version. This is a *write-write conflict*. We follow the first-writer-wins rule and force transaction T to abort.

Suppose T finds that V is updatable. It then creates the new version and proceeds to install it in the database. The first step is to atomically store T's transaction ID in V's End field to prevent other transactions from updating V. If the store fails because the End field has changed, T must abort because some other transaction has sneaked in and updated V before T managed to install its update. If the store succeeds, T then connects the new version into all indexes it participates in. T also saves pointers to the old and new versions; they will be needed during postprocessing.

## 2.7 Commit Dependencies

A transaction T1 has a commit dependency on another transaction T2, if T1 is allowed to commit only if T2 commits. If T2 aborts, T1 must also abort, so cascading aborts are possible. T1 acquires a commit dependency either by speculatively reading or speculatively ignoring a version, instead of waiting for T2 to commit..

We implement commit dependencies by a register-and-report approach: T1 registers its dependency with T2 and T2 informs T1 when it has committed or aborted. Each transaction T contains a counter, CommitDepCounter, that counts how many unresolved commit dependencies it still has. A transaction cannot commit until this counter is zero. In addition, T has a Boolean variable AbortNow that other transactions can set to tell T to abort. Each transaction T also has a set, CommitDepSet, that stores transaction IDs of the transactions that depend on T.

To take a commit dependency on a transaction T2, T1 increments its CommitDepCounter and adds its transaction ID to T2's CommitDepSet. When T2 has committed, it locates each transaction in its CommitDepSet and decrements their CommitDepCounter. If T2 aborted, it tells the dependent transactions to also abort by setting their AbortNow flags. If a dependent transaction is not found, this means that it has already aborted.

Note that a transaction with commit dependencies may not have to wait at all - the dependencies may have been resolved before it is ready to commit. Commit dependencies consolidate all waits into a single wait and postpone the wait to just before commit.

Some transactions may have to wait before commit. Waiting raises a concern of deadlocks. However, deadlocks cannot occur because an older transaction never waits on a younger transaction. In a wait-for graph the direction of edges would always be from a younger transaction (higher end timestamp) to an older transaction (lower end timestamp) so cycles are impossible.

## 3. OPTIMISTIC TRANSACTIONS

This section describes in more detail the processing performed in the different phases for optimistic transactions. We first consider serializable transactions and then discuss lower isolation levels.

The original paper by Kung and Robinson [17] introduced two validation methods: backward validation and forward validation. We use backward validation but optimize it for in-memory storage. Instead of validating a read set against the write sets of all other transactions, we simply check whether a version that was read is still visible as of the end of the transaction. A separate write phase is not needed; a transaction's updates become visible to other transactions when the transaction changes its state to Committed.

A serializable optimistic transaction keeps track of its reads, scans and writes. To this end, a transaction object contains three sets:

1. **ReadSet** contains pointers to every version read;
2. **ScanSet** stores information needed to repeat every scan;
3. **WriteSet** contains pointers to versions updated (old and new), versions deleted (old) and versions inserted (new).

## 3.1 Normal Processing Phase

Normal processing consists of scanning indexes (see Section 2.1) to locate versions to read, update, or delete. Insertion of an entirely new record or updating an existing record creates a new version that is added to all indexes for records of that type.

To do an index scan, a transaction T specifies an index $I$, a predicate $P$, and a logical read time RT. The predicate is a conjunction $P = P_s \wedge P_r$ where $P_s$ is a search predicate that determines what part of the index to scan and $P_r$ is an optional residual predicate. For a hash index, $P_s$ is an equality predicate on columns of the hash key. For an ordered index, $P_s$ is a range predicate on the ordering key of the index.

We now outline the processing during a scan when T runs at serializable isolation level. All reads specify T's begin time as the logical read time.

**Start scan.** When a scan starts, it is registered in T's ScanSet so T can check for phantoms during validation. Sufficient information must be recorded so that the scan can be repeated. This includes at least the following: index I and predicates $P_s$ and $P_r$. Exactly how the predicates are specified depends on the implementation.

**Check predicate.** If a version V doesn't satisfy $P$, it is ignored and the scan proceeds. If the scan is a range scan and the index key exceeds the upper bound of the range, the scan is terminated.

**Check visibility.** Next we check whether version V is visible to transaction T as of time RT (section 2.5). The result of the test may be conditional on another transaction T2 committing. If so, T registers a commit dependency with T2 (section 2.7). If the visibility test returns false, the scan proceeds to the next version.

**Read version.** If T intends to just read V, no further checks are required. T records the read by adding a pointer to V to its ReadSet. The pointer will be used during validation.

**Check updatability.** If transaction T intends to update or delete V, we must check whether the version is updatable. A visible version is updatable if its End field equals infinity or it contains a transaction ID and the referenced transaction has aborted. Note that speculative updates are allowed, that is, an uncommitted version can be updated but the transaction that created it must have completed normal processing.

**Update version.** To update V, T first creates a new version VN and then atomically sets V's End field to T's transaction ID. This fails if some other transaction T2 has already set V's End field. This is a write-write conflict and T must abort.

If T succeeds in setting V's End field (the most likely outcome), this serves as an exclusive write lock on V because it prevents further updates of V. T records the update by adding two pointers to its WriteSet: a pointer to V (old version) and a pointer to VN (new version). These pointers are used later for multiple purposes: for logging new versions during commit, for postprocessing after commit or abort, and for locating old versions when they are no longer needed and can be garbage collected.

The new version VN is not visible to any other transaction until T precommits, therefore T can proceed to include VN in all indexes that the table participates in.

**Delete version.** A delete is an update of V that doesn't create a new version. The end timestamp of V is first checked and then set in the same way as for updates. If this succeeds, a pointer to V (old version) is added to the write set and the delete is complete.

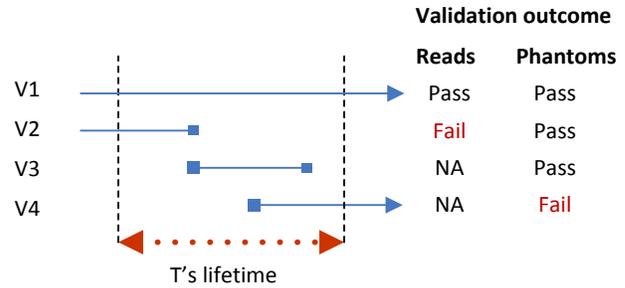

Figure 3: Possible validation outcomes.

When transaction T reaches the end of normal processing, it precommits and begins its preparation phase. Precommit simply consists of acquiring the transaction's end timestamp and setting the transaction state to Preparing.

## 3.2 Preparation Phase

The preparation phase of an optimistic transaction consists of three steps: read validation, waiting for commit dependencies, and logging. We discuss each step in turn.

Validation consists of two steps: checking visibility of the versions read and checking for phantoms. To check visibility transaction T scans its ReadSet and for each version read, checks whether the version is still visible as of the end of the transaction. To check for phantoms, T walks its ScanSet and repeats each scan looking for versions that came into existence during T's lifetime and are visible as of the end of the transaction. (T may acquire additional commit dependencies during validation but only if it speculatively ignores a version.)

Figure 3 illustrates the different cases that can occur. It shows the lifetime of a transaction T, the valid times of versions V1 to V4 of four different records, and the expected outcome of read validation and phantom detection. We assume that all four versions satisfy the search predicate used by T and that they were all created and terminated by transactions other than T.

Version V1 is visible to T both at its start and end. If V1 is included in T's ReadSet, it passes read validation and also phantom detection.

Version V2 is visible to T as of its start timestamp but not at the end of the transaction. If V2 is included in T's ReadSet, it fails read validation. V2 is not a phantom.

Version V3 both began and ended during T's lifetime, so it is not visible to T at the start or at the end of the transaction. It is not included in T's ReadSet so it won't be subject to read validation. V3 is not visible at the end of T, so V3 is not a phantom.

Version V4 was created during T's lifetime and is visible at the end of T, so V4 is a phantom. It is not included in T's ReadSet because it was not visible as of T's start time.

If T fails validation, it is not serializable and must abort. If T passes validation, it must wait for outstanding commit dependencies to be resolved, if it has any. More specifically, T can proceed to the postprocessing phase if either its CommitDepCounter is zero or its AbortNow flag is set.

To complete the commit, T scans its WriteSet and writes the new versions it created to a persistent log. Commit ordering is determined by transaction end timestamps, which are included in the log records, so multiple log streams on different devices can be used. Deletes are logged by writing a unique key or, in the worst case, all columns of the deleted version. After the log writes have been completed, T sets its transaction state to Committed, thereby signaling that the commit is done.

## 3.3 Postprocessing

During this phase a committed transaction TC propagates its end timestamp to the Begin and End fields of new and old versions, respectively, listed in its WriteSet. An aborted transaction TA sets the Begin field of its new versions to infinity, thereby making them invisible to all transactions, and attempts to reset the End fields of its old versions to infinity. However, another transaction may already have detected the abort, created another new version and reset the End field of the old version. If so, TA leaves the End field value unchanged.

The transaction then processes all outgoing commit dependencies listed in its CommitDepSet. If it aborted, it forces the dependent transactions to also abort by setting their AbortNow flags. If it committed, it decrements the target transaction's CommitDep­Counter and wakes up the transaction if the count becomes zero.

Once postprocessing is done, other transactions no longer need to refer to the transaction object. It can be removed from the transaction table but it will not be discarded entirely; the pointers to old versions in its WriteSet are needed for garbage collection.

## 3.4 Lower Isolation Levels

Enforcing lower isolation levels is cheaper. A transaction requiring a higher isolation level bears the full cost of enforcing the higher level and does not penalize "innocent bystanders".

**Repeatable read:** Repeatable read is required to enforce read stability but not to prevent phantoms. We implement repeatable read simply by validating a transaction's ReadSet before commit. As phantom detection is not required, a transaction's scans are not recorded. Transaction begin time is used as the logical read time.

**Read committed:** Read committed only guarantees that all versions read are committed. We implement read committed by always using the current time as the logical read time. No validation is required so a transaction's reads and scans are not recorded.

**Snapshot isolation:** Implementing snapshot isolation is straightforward in our case: always read as of the beginning of the transaction. No validation is needed so scans and reads are not tracked.

**Read-only transactions:** If a transaction is known to be read-only, the best performance is obtained by running it under snapshot isolation or read committed depending on whether it needs a transaction-consistent view or not.

## 4. PESSIMISTIC TRANSACTIONS

A pessimistic transaction prevents its reads from being invalidated by acquiring read locks. This section describes our design of multiversion locking optimized for main-memory databases

A serializable pessimistic transaction must keep track of which versions it read, which hash buckets it scanned, and its new and old versions. To this end, the transaction maintains three sets:

1. **ReadSet** contains pointers to versions read locked by the transaction;
2. **BucketLockSet** contains pointers to hash buckets visited and locked by the transaction;
3. **WriteSet** contains references to versions updated (old and new), versions deleted (old) and versions inserted (new).

## 4.1 Lock Types

We use two types of locks: record locks and bucket locks. Record locks are placed on versions to ensure read stability. Bucket locks are placed on (hash) buckets to prevent phantoms. The name reflects their use for hash indexes in our prototype but range locks for ordered indexes can be implemented in the same way.

### 4.1.1 Record Locks

Updates or deletes can only be applied to the latest version of a record; older versions cannot be further updated. Thus, locks are required only for the latest version of a record; never for older versions. So what's needed is an efficient many-readers-single-writer lock for this case.

We do not want to store record locks in a separate table – it's too slow. Instead we embed record locks in the End field of versions so no extra space is required. In our prototype, the End field of a version is 64 bits. As described earlier, this field stores either a timestamp or a transaction ID with one bit indicating what the field contains. We change how we use this field to make room for a record lock.

1. ContentType (1 bit): indicates the content type of the remaining 63 bits
2. Timestamp (63 bits): when ContentType is zero.
3. RecordLock (63 bits): when ContentType is one.
   3.1. NoMoreReadLocks (1 bit): a flag set when no further read locks are accepted. Used to prevent starvation.
   3.2. ReadLockCount (8 bits): number of read locks.
   3.3. WriteLock (54 bits): ID of the transaction holding a write lock on this version or infinity (max value).

We do not explicitly keep track of which transactions have a version read locked. Each transaction records its ReadSet so we can find out by checking the ReadSets of all current transactions. This is only needed for deadlock detection which occurs infrequently.

A transaction acquires a read lock on a version V by atomically incrementing V's ReadLockCount. No further read locks can be acquired if the counter has reached its max value (255) or the NoMoreReadlocks flag is set. If so, the transaction aborts.

A transaction write locks a version V by atomically copying its transaction ID into the WriteLock field. This action both write locks the version and identifies who holds the write lock.

### 4.1.2 Bucket Locks (Range Locks)

Bucket locks are used only by serializable transactions to prevent phantoms. When a transaction TS begins a scan of a hash bucket, it locks the bucket. Multiple transactions can have a bucket locked. A bucket lock consists of the following fields.

1. **LockCount**: number of locks on this bucket.
2. **LockList**: list of (serializable) transactions holding a lock on this bucket.

The current implementation stores the LockCount in the hash bucket to be able to check quickly whether the bucket is locked. LockLists are implemented as arrays stored in a separate hash table with the bucket address as the key.

To acquire a lock on a bucket B, a transaction TS increments B's LockCount, locates B's LockList, and adds its transaction Id to the list. To release the lock it deletes its transaction ID from B's LockList and decrements the LockCount.

Range locks in an ordered index can be implemented in the same way. If the index is implemented by a tree structure, a lock on a node locks the subtree rooted at that node. If the index is implemented by skip lists, locking a tower locks the range from that tower to the next tower of the same height.

## 4.2 Eager Updates, Wait-For Dependencies

In a traditional implementation of multiversion locking, an update transaction TU would block if it attempts to update or delete a read locked version or attempts to insert or update a version in a locked bucket. This may lead to frequent blocking and thread switching. A thread switch is expensive, costing several thousand instructions. In a main-memory system, just a few thread switches can add significantly to the cost of executing a transaction.

To avoid blocking we allow a transaction TU to eagerly update or delete a read locked version V *but,* to ensure correction serialization order, TU cannot precommit until all read locks on V have been released. Similarly, a transaction TR can acquire a read lock on a version that is already write locked by another transaction TU. If so, TU cannot precommit until TR has released its lock.

Note that an eager update or delete is not speculative because it doesn't matter whether TR commits or aborts; it just has to complete and release its read lock.

The same applies to locked buckets. Suppose a bucket B is locked by two (serializable) transactions TS1 and TS2. An update transaction TU is allowed to insert a new version into B but it is not allowed to precommit before TS1 and TS2 have completed and released their bucket locks.

We enforce correct serialization order by *wait-for dependencies*. A wait-for dependency forces an update transaction TU to wait before it can acquire an end timestamp and begin commit processing. There are two flavors of wait-for dependencies, *read lock dependencies* and *bucket lock dependencies* that differ in what event they wait on.

A transaction T needs to keep track of both incoming and outgoing wait-for dependencies. T has an incoming dependency if it waits on some other transaction and an outgoing dependency if some other transaction waits on it. To track wait-for dependencies, the following fields are included in each transaction object.

For incoming wait-for dependencies:
- **WaitForCounter**: indicates how many incoming dependencies the transaction is waiting for.
- **NoMoreWaitFors**: when set the transaction does not allow additional incoming dependencies. Used to prevent starvation by incoming dependencies continuously being added.

For outgoing wait-for dependencies:
- **WaitingTxnList**: IDs of transactions waiting on this transaction to complete.

### 4.2.1 Read Lock Dependencies.

A transaction TU that updated or deleted a version V has a wait-for dependency on V as long as V is read locked. TU is not allowed to acquire an end timestamp and begin commit processing unless V's ReadLockCount is zero.

When a transaction TU updates or deletes a version V, it acquires a write lock on V by copying its transaction ID into the WriteLock field. If V's ReadLockCount is greater than zero, TU takes a wait-for dependency on V simply by incrementing its WaitForCounter.

TU may also acquire a wait-for dependency on V by another transaction TR taking a read lock on V. A transaction TR that wants to read a version V must first acquire a read lock on V by incrementing V's ReadLockCount. If V's NoMoreReadLocks flag is set or ReadLockCount is at max already, lock acquisition fails and TR aborts. Otherwise, if V is not write locked or V's ReadLockCount is greater than zero, TR increments V's ReadLockCount and proceeds. However, if V is write locked by a transaction TU and this is the first read lock on V (V's ReadLockCount is zero), TR must force TU to wait on V. TR checks TU's NoMoreWaitFors flag. If it is set, TU cannot install the wait-for dependency and aborts. Otherwise everything is in order and TR acquires the read lock by incrementing Vs' ReadLockCounter and installs the wait-for dependency by incrementing TU's WaitForCounter.

When a transaction TR releases a read lock on a version V, it may also need to release a wait-for dependency. If V is not write locked, TR simply decrements V's ReadLockCounter and proceeds. The same applies if V is write locked and V's ReadLockCounter is greater than one. However, if V is write locked by a transaction TU and V's ReadLockCounter is one, TR is about to release the last read lock on V and therefore must also release TU's wait-for dependency on V. TR atomically sets V's ReadLockCounter to zero and V's NoMoreReadLocks to true. If this succeeds, TR locates TU and decrements TU's WaitForCounter.

Setting the NoMoreReadLocks flag before releasing the wait-for dependency is necessary because this may be TU's last wait-for dependency. If so, TU is free to acquire an end timestamp and being its commit processing. In that case, TU's commit cannot be further postponed by taking out a read lock on V. In other words, further read locks on V would have no effect.

### 4.2.2 Bucket Lock Dependencies.

A serializable transaction TS acquires a lock on a bucket B by incrementing B's LockCounter and adding its transaction ID to B's LockList. The purpose of TR's bucket lock is not to disallow new versions from being added to B but to prevent them from becoming visible to TR. That is, another transaction TU can add a version to B but, if it does, then TU cannot precommit until TS has completed its processing and released its lock on B. This is enforced by TU obtaining a wait-for dependency on TS.

TU can acquire this type of dependency either by acquiring one itself or by having one imposed by TS. We discuss each case.

Suppose that, as a result of an update or insert, TU is about to add a new version V to a bucket B. TU checks whether B has any bucket locks. If it does, TU takes out a wait-for dependency on every transaction TS in B's LockList by adding its own transaction ID to TS's WaitForList and incrementing its own WaitForCounter. If TU's NoMoreWaitFors flag is set, TU can't take out the dependency and aborts.

Suppose a serializable transaction TS is scanning a bucket B and encounters a version V that satisfies TS's search predicate but the version is not visible to TS, that is, V is write locked by a transaction TU that is still active. If TU commits before TS, V becomes a phantom to TS. To prevent this from happening, TS registers a wait-for dependency on TU's behalf by adding TU's transaction ID to its own WaitingTxnList and incrementing TU's WaitForCounter. If TU's NoMoreWaitFors flag is set, TS can't impose the dependency and aborts.

When a serializable transaction TS has precommitted and acquired its end timestamp, it releases its outgoing wait-for dependencies. It scans its WatingTxnList and, for each transaction T found, decrements T's WaitForCounter.

## 4.3 Processing Phases

This section describes how locking affects the processing done in the different phases of a transaction.

### 4.3.1 Normal Processing Phase

Recall that normal processing consists of scanning indexes to select record versions to read, update, or delete. An insertion or update creates a new version that has to be added to all indexes for records of that type.

We now outline what a pessimistic transaction T does differently than an optimistic transaction during a scan and how this depends on T's isolation level. For snapshot isolation, the logical read time is always the transaction begin time. For all other isolation levels, it is the current time which has the effect that the read sees the latest version of a record.

**Start scan.** If T is a serializable transaction, it takes out a bucket lock on B to prevent phantoms and records the lock in its BucketLockSet. Other isolation levels do not take out a bucket lock.

**Check predicate.** Same as for optimistic transactions.

**Check visibility.** This is done in the same way as for optimistic transaction, including taking out commit dependencies as needed. If a version V is not visible, it is ignored and the scan continues for all isolations levels except serializable. If T is serializable and V is write locked by a transaction TU that is still active, V is a potential phantom so T forces TU to delay its precommit by imposing a wait-for dependency on TU.

**Read version**. If T runs under serializable or repeatable read and V is a latest version, T attempts to acquire a read lock on V. If T can't acquire the read lock, it aborts. If T runs under a lower isolation level or V is not a latest version, no read lock is required.

**Check updatability.** The same as for optimistic transactions.

**Update version**. As for optimistic transactions, T creates a new version N, sets V's WriteLock and, if V was read locked, takes out a wait-for dependency on V by incrementing its own WaitForCounter. T then proceeds to add N to all indexes it participates in. If T adds N to a locked index bucket B, it takes out wait-for dependencies on all (serializable) transactions holding locks on B.

**Delete version.** A delete is essentially an update of V that doesn't create a new version. T sets V's WriteLock and if V was read locked, takes out a wait-for dependency on V by incrementing its own WaitForCounter.

When transaction T reaches the end of normal processing, it releases its read locks and its bucket locks, if any. If it has outstanding wait-for dependencies (its WaitForCounter is greater than zero), it waits. Once its WaitForCounter is zero, T precommits, that is, acquires an end timestamp and sets its state to Validating.

### 4.3.2 Preparation Phase

Pessimistic transactions require no validation – that's taken care of by locks. However, a pessimistic transaction T may still have outstanding commit dependencies when reaching this point. If so, T waits until they have been resolved and then proceeds to write to the log and commit. If a commit dependency fails, T aborts.

### 4.3.3 Postprocessing Phase

Postprocessing is the same as for optimistic transactions. Note that there is no need to explicitly release write locks; this is automatically done when the transaction updates Begin and End fields.

## 4.4 Deadlock Detection

Commit dependencies are only taken on transactions that have already precommitted and are completing validation. As discussed earlier (Section 2.7) commit dependencies cannot cause or be involved in a deadlock.

Wait-for dependencies, however, can cause deadlocks. To detect deadlocks we build a standard wait-for graph by analyzing the wait-for dependencies of all transactions that are currently blocked. Once the wait-for graph has been built, any algorithm for finding cycles in graphs can be used. Our prototype uses Tarjan's algorithm [25] for finding strongly connected components.

A wait-for graph is a directed graph with transactions as nodes and waits-for relationships as edges. There is an edge from transaction T2 to transaction T1 if T2 is waiting for T1 to complete. The graph is constructed in three steps.

1. **Create nodes.** Scan the transaction table and for each transaction T found, create a node N(T) if T has completed its normal processing and is blocked because of wait-for dependencies
2. **Create edges from explicit dependencies.** Wait-for dependencies caused by bucket locks are represented explicitly in WaitingTxnLists. For each transaction T1 in the graph and each transaction T2 in T1's WaitingTxnList, add an edge from T2 to T1.
3. **Create edges from implicit dependencies.** A wait-for dependency on a read-locked version V is an implicit representation of wait-for dependencies on all transaction holding read locks on V. We can find out which transactions hold read locks on V by checking transaction read sets. Edges from implicit dependencies can be constructed as follows. For each transaction T1 in the graph and each version V in T1's ReadLockSet: if V is write locked by a transaction T2 and T2 is in the graph, add an edge from T2 to T1.

While the graph is being constructed normal processing continues so wait-for dependencies may be created and resolved and transactions may become blocked or unblocked. Hence, the final graph obtained may be imprecise, that is, it may differ from the graph that would be obtained if normal processing stopped. But this doesn't matter because if there truly is a deadlock neither the nodes nor the edges involved in the deadlock can disappear while the graph is being constructed. There is a small chance of detecting a false deadlock but this is handled by verifying that the transactions participating in the deadlock are still blocked and the edges are still covered by unresolved wait-for dependencies.

## 4.5 Peaceful Coexistence

An interesting property of our design is that optimistic and pessimistic transactions can be mixed. The change required to allow optimistic transactions to coexist with pessimistic transactions is straightforward: optimistic update transactions must honor read locks and bucket locks. Making an optimistic transaction T honor read locks and bucket locks requires the following changes:

1. When T write locks a version V, it uses only a 54-bit transaction ID and doesn't overwrite read locks.
2. When T updates or deletes a version V that is read locked, it takes a wait-for dependency on V.
3. When T inserts a new version into a bucket B, it checks for bucket locks and takes out wait-for dependencies as needed.

# 5. EXPERIMENTAL RESULTS

Our prototype storage engine implements both the optimistic and the pessimistic scheme. We also implemented a single-version storage engine with locking for concurrency control. The implementation is optimized for main-memory databases and does not use a central lock manager, as this can become a bottleneck [19]. Instead, we embed a lock table in every index and assign each hash key to a lock in this partitioned lock table. A lock covers all records with the same hash key which automatically protects against phantoms. We use timeouts to detect and break deadlocks.

The experiments were run on a two-socket Intel Xeon X5650 @ 2.67 GHz (Nehalem) that has six cores per socket. Hyper-Threading was enabled. The system has 48 GB of memory, 12 MB L3 cache per socket, 256 KB L2 cache per core, and two separate 32 KB L1-I and L1-D caches per core. This is a NUMA system and memory latency is asymmetric: accessing memory on the remote socket is approximately 30% slower than accessing local memory. We size hash tables appropriately so there are no collisions. The operating system is Windows Server 2008 R2.

Each transaction generates log records but these are asynchronously written to durable storage; transactions do not wait for log I/O to complete. Asynchronous logging allows us to submit log records in batches (group commit), greatly reducing the number of I/O operations. The I/O bandwidth required is also moderate: Each update produces a log record that stores the difference between the old and new versions, plus 8 bytes of metadata. Even with millions of updates per second, the I/O bandwidth required is within what even a single desktop hard drive can deliver. This choice allows us to focus on the effect of concurrency control.

Traditional disk-based transaction processing systems require hundreds of concurrently active transactions to achieve maximum throughput. This is to give the system useful work to do while waiting for disk I/O. Our main-memory engine does not wait for disk I/O, so there is no need to overprovision threads. We observed that the CPU is fully utilized when the multi-programming level equals the number of hardware threads; allowing more concurrent transactions causes throughput to drop. We therefore limited the number of concurrently active transactions to be at most 24, which matches the number of threads our machine supports.

We experiment with three CC schemes: the single-version locking engine (labeled "1V"), the multi-version engine when transactions run optimistically ("MV/O") and the multi-version engine where transactions run pessimistically ("MV/L").

## 5.1 Homogeneous Workload

We first show results from a parameterized artificial workload. By varying the workload parameters we can systematically explore how sensitive the different schemes are to workload characteristics. We focus on short update transactions which are common for OLTP workloads. The workload consists of a single transaction type that performs $R$ reads and $W$ writes against a table of $N$ records with a unique key. Each row is 24 bytes, and reads and writes are uniformly and randomly scattered over the $N$ records.

The memory footprint of the database differs for each concurrency control scheme. In 1V, each row consumes 24 bytes (payload) plus 8 bytes for the "next" pointer of the hash table. Both MV schemes additionally use 16 bytes to store the Begin and End fields (cf. Figure 1), but the total consumption depends on the average number of versions required by the workload. Comparing the two MV schemes, MV/L has the biggest memory footprint, due to the additional overhead of maintaining a bucket lock table.

### 5.1.1 Scalability (Read Committed)

We first show how transaction throughput scales with increasing multiprogramming level. For this experiment each transaction performs 10 reads and 2 writes ($R=10$ and $W=2$) against a table with $N=10,000,000$ rows. The isolation level is Read Committed.

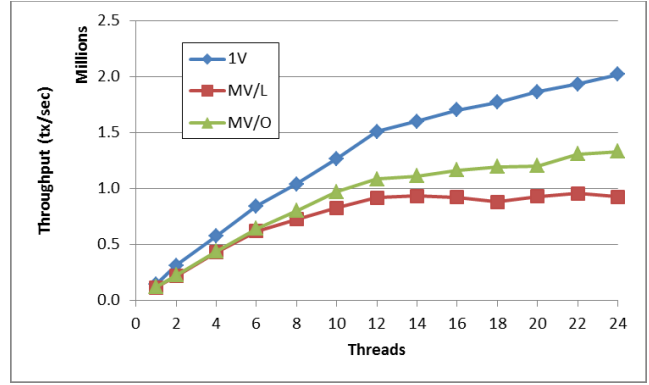

**Figure 4: Scalability under low contention**

Figure 4 plots transaction throughput (y-axis) as the multiprogramming level increases (x-axis). Under low contention, throughput for all three schemes scales linearly up to six threads. After six threads, we see the effect of the higher access latency as the data is spread among two NUMA nodes, and beyond twelve threads we see the effect of HyperThreading.

For the 1V scheme, HyperThreading causes the speed-up rate to drop but the system still continues to scale linearly, reaching a maximum of over 2M transactions/sec. The multiversion schemes have lower throughput because of the overhead of version management and garbage collection. Creating a new version for every update and cleaning out old versions that are no longer needed is obviously more expensive than updating in place.

Comparing the two multiversion schemes, MV/L has 30% lower performance than MV/O. This is caused by extra writes for tracking dependencies and locks, which cause increased memory traffic. It takes MV/L 20% more cycles to execute the same number of instructions and the additional control logic translates into 10% more instructions per transaction.

### 5.1.2 Scaling under Contention (Read Committed)

Records that are updated very frequently (hotspots) pose a problem for all CC schemes. In locking schemes, high contention causes transactions to wait because of lock conflicts and deadlocks. In optimistic schemes, hotspots result in validation failures and write-write conflicts, causing high abort rates and wasted work. At the hardware level, some data items are accessed so frequently that they practically reside in the private L1 or L2 caches of each core. This stresses the hardware to the limits, as it triggers very high core-to-core traffic to keep the caches coherent.

We simulate a hotspot by running the same $R=10$ and $W=2$ transaction workload from Section 5.1.1 on a very small table with just $N=1,000$ rows. Transactions run under Read Committed. Figure 5 shows the throughput under high contention. Even in this admittedly extreme scenario, all schemes achieve a throughput of over a million transactions per second, with MV/O slightly ahead of both locking schemes. 1V achieves its highest throughput at six threads, then drops and stays flat after 8 threads.

306

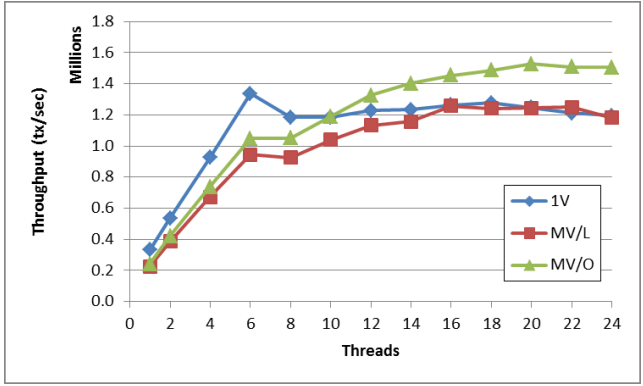

Figure 5: Scalability under high contention

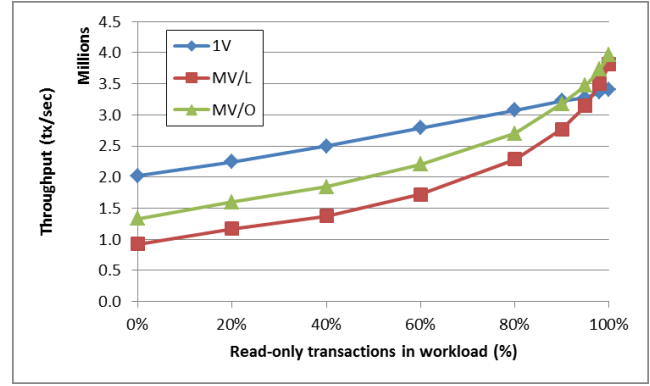

Figure 6: Impact of read-only transactions (low contention)

### 5.1.3 Higher IsolationLlevels

The experiments in the previous section ran under Read Committed isolation level, which is the default isolation level in many commercial database engines [23], as it prevents dirty reads and offers high concurrency. Higher isolation levels prevent more anomalies but reduce throughput. In this experiment, we use the same workload from Section 5.1.1, we fix the multiprogramming level to 24 and we change the isolation level.

|  | Read Committed | Repeatable Read |  | Serializable |  |
|---|---|---|---|---|---|
|  | tx/sec | tx/sec | % drop vs RC | tx/sec | % drop vs RC |
| **1V** | 2,080,492 | 2,042,540 | 1.8% | 2,042,571 | 1.8% |
| **MV/L** | 974,512 | 963,042 | 1.2% | 877,338 | 10.0% |
| **MV/O** | 1,387,140 | 1,272,289 | 8.3% | 1,120,722 | 19.2% |

**Table 3: Throughput at higher isolation levels, and percentage drop compared to Read Committed (RC)**

In Table 3, we report the transaction throughput from each scheme and isolation level. We also report the throughput drop as a percentage of the throughput when running under the Read Committed isolation level.

The overhead for Repeatable Read for both locking schemes is very small, less than 2%. MV/O needs to repeat the reads at the end of the transaction, and this causes an 8% drop in throughput. For Serializable, the 1V scheme protects the hash key with a lock, and this guarantees phantom protection with very low overhead (2%). Both MV schemes achieve serializability at a cost of 10%–19% lower throughput: MV/L acquires read locks and bucket locks, while MV/O has to repeat each scan during validation. Under MV/O, however, a transaction requesting a higher isolation level bears the full cost of enforcing the higher isolation. This is not the case for locking schemes.

## 5.2 Heterogeneous Workload

The workload used in the previous section represents an extreme update-heavy scenario. In this section we fix the multiprogramming level to 24 and we explore the impact of adding read-only transactions in the workload mix.

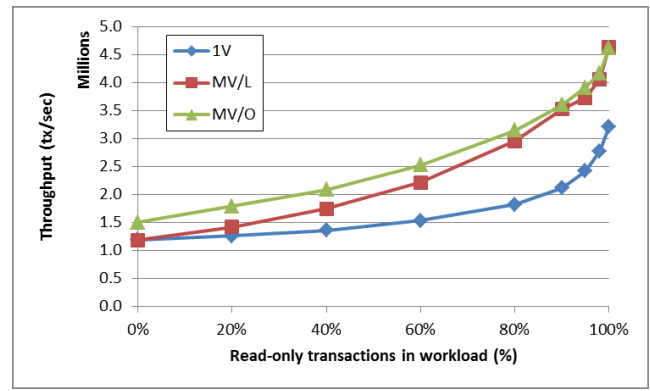

Figure 7: Impact of read-only transactions (high contention)

### 5.2.1 Impact of Short Read Transactions

In this experiment we change the ratio of read and update transactions. There are two transaction types running under Read Committed isolation: the update transaction performs 10 reads and 2 writes ($R=10$ and $W=2$), while the read-only transaction performs 10 reads ($R=10$ and $W=0$).

Figure 6 shows throughput (y-axis) as the ratio of read-only transactions varies in the workload (x-axis) in a table with *10,000,000* rows. The leftmost point (x=0%) reflects the performance of the update-only workload of Section 5.1.1 at 24 threads. As we add read-only transactions to the mix, the performance gap between all schemes closes. This is primarily because the update activity is reduced, reducing the overhead of garbage collection.

The MV schemes outperform 1V when most transactions are read-only. When a transaction is read-only, the two MV schemes behave identically: transactions read a consistent snapshot and do not need to do any locking or validation. In comparison, 1V has to acquire and release short read locks for cursor stability even for read-only transactions which impacts performance.

In Figure 7 we repeat the same experiment but simulate a hotspot by using a table of *1,000* rows. The leftmost point (x=0%) again reflects the performance of the update-only workload of Section 5.1.12 under high contention at 24 threads. The MVCC schemes have a clear advantage at high contention, as snapshot isolation prevents read-only transactions from interfering with writers. When 80% of the transactions are read-only, the MVCC schemes achieve 63% and 73% higher throughput than 1V.

## 5.2.2 Impact of Long Read Transactions

Not all transactions are short in OLTP systems. Users often need to run operational reporting queries on the live system. These are long read-only transactions that may touch a substantial part of the database. The presence of a few long-running queries should not severely affect the throughput of "normal" OLTP transactions.

This experiment investigates how the three concurrency control methods perform in this scenario. We use a table with *10,000,000* rows and fix the number of concurrently active transactions to be 24. The workload consists of two transaction types: (a) Long, transactionally consistent (Serializable), read-only queries that touch 10% of the table (*R=1,000,000* and *W=0*) and (b) Short update transactions with *R=10* and *W=2*.

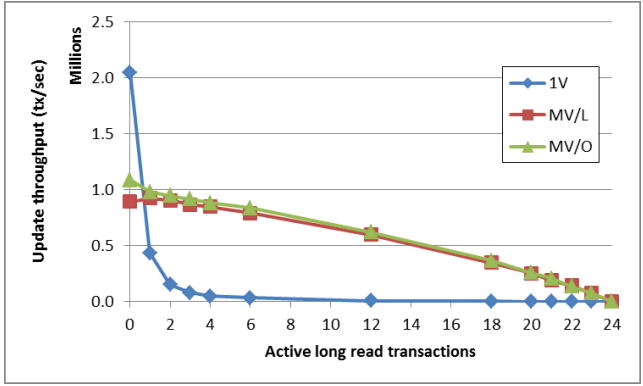

**Figure 8: Update throughput with long read transactions**

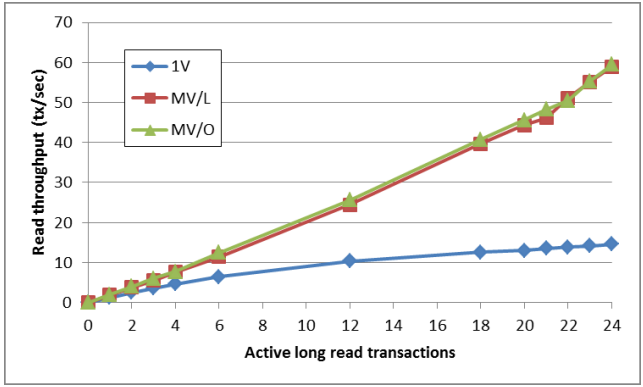

**Figure 9: Read throughput with long read transactions**

Figures 8 and 9 show update and read throughput (y-axis) as we vary the number of concurrent long read-only transactions in the system (x-axis). At the leftmost point (x=0) all transactions are short update transactions, while at the rightmost point (x=24) all transactions are read-only. At x=6, for example, there are 6 read-only and 24-6=18 short update transactions running concurrently.

Looking at the update throughput in Figure 7, we can see that 1V is twice as fast as the MV schemes when all transactions are short update transactions, at x=0. (This is consistent with our findings from the experiments with the homogeneous workload in Section 5.1.) However the picture changes completely once a single long read-only transaction is present in the system. At x=1, update throughput drops 75% for the single version engine. In contrast, update throughput drops only 5% for the MV schemes, making MV twice as fast as 1V. The performance gap grows as we allow more read-only transactions. When 50% of the active transactions are long readers, at x=12, MV has 80X higher update throughput than 1V. In terms of read throughput (Figure 8), both MV schemes consistently outperform 1V.

## 5.3 TATP Results

The workloads used in the previous sections allowed us to highlight fundamental differences between the three concurrency control mechanisms. However, real applications have higher demands than randomly reading and updating values in a single index. We conclude our experimental evaluation by running a benchmark that models a simple but realistic OLTP application.

The TATP benchmark [24] simulates a telecommunications application. The database consists of four tables with two indexes on each table to speed up lookups. The benchmark runs a random mix of seven short transactions; each transaction performs less than 5 operations on average. 80% of the transactions executed only query the database, while 16% update, 2% insert, and 2% delete items. We sized the database for 20 million subscribers and generated keys using the non-uniform distribution that is specified in the benchmark. All transactions run under Read Committed.

|  | 1V | MV/L | MV/O |
|---|---|---|---|
| **Transactions per second** | 4,220,119 | 3,129,816 | 3,121,494 |

**Table 4: TATP results**

Table 4 shows the number of committed transactions per second for each scheme. Our concurrency control mechanisms can sustain a throughput of several millions of transaction per second on a low-end server machine. This is an order of magnitude higher than previously published TATP numbers for disk-based systems [19] or main memory systems [14].

## 6. RELATED WORK

Concurrency control has a long and rich history going back to the beginning of database systems. Several excellent surveys and books on concurrency control are available [4], [16], [26].

Multiversion concurrency control methods also have a long history. Chapter 5 in [4] describes three multiversioning methods: multiversion timestamp ordering (MVTO), two-version two-phase locking (2V2PL), and a multiversion mixed method. 2V2PL uses at most two versions: last committed and updated uncommitted. They also sketch a generalization that allows multiple uncommitted versions and readers are allowed to read uncommitted versions. The mixed method uses MVTO for read-only transactions and Strict 2PL for update transactions.

The optimistic approach to concurrency control originated with Kung and Robinson, but they only considered single-version databases [17]. Many multiversion concurrency control schemes have been proposed [2], [5], [6], [8], [9], [13], [18], [20], but we are aware of only two that take an optimistic approach: Multiversion Serial Validation (MVSV) by Carey [11], [12] and Multiversion Parallel Validation (MVPV) by Agrawal et al [1]. While the two schemes are optimistic and multiversion, they differ significant from our scheme. Their isolation level is repeatable read; other isolation levels are not discussed. MVSV does validation serially so validation quickly becomes a bottleneck. MVPV does validation in parallel but installing updates after validation is done serially. In comparison, the only critical section in our method is acquiring timestamps; everything else is done in parallel. Acquiring a timestamp is a single instruction (an atomic increment) so the critical section is extremely short.

Snapshot isolation (SI) [3] is a multiversioning scheme used by many database systems. Several commercial database systems support snapshot isolation to isolate read-only transactions from updaters: Oracle [22], Postgres [21] and SQL Server [23] and possibly others. However, SI is not serializable and many papers have considered under what circumstances SI is serializable or how to make it serializable. Cahill et al published a complete and practical solution in 2008 [9]. Their technique requires that transactions check for read-write dependencies. Their implementation uses a standard lock manager and transactions acquire "locks" and check for read-write dependencies on every read and write. The "locks" are non-blocking and used only to detect read-write dependencies. Whether their approach can be implemented efficiently for a main-memory DBMS is an open question. Techniques such as validating by checking repeatability of reads and predicates have already been used in the past [7].

Oracle TimesTen [22] and IBM's solidDB [10] are two commercially available main-memory DBMSs. TimesTen uses single-version locking with multiple lock types (shared, exclusive, update) and multiple granularities (row, table, database). For main-memory tables, solidDB also uses single-version locking with multiple lock types (shared, exclusive, update) and two granularities (row, table). For disk-based tables, solidDB supports both optimistic and pessimistic concurrency control. .

## 7. CONCLUDING REMARKS

In this paper we investigated concurrency control mechanisms optimized for main memory databases. The known shortcomings of traditional locking led us to consider solutions based on multi-versioning. We designed and implemented two MVCC methods, one optimistic using validation and one pessimistic using locking. For comparison purposes we also implemented a variant of single-version locking optimized for main memory databases. We then experimentally evaluated the performance of the three methods. Several conclusions can be drawn from the experimental results.

- Single-version locking can be implemented efficiently and without lock acquisition becoming a bottleneck.
- Single-version locking is fragile; it performs well when transactions are short and contention is low but suffers under more demanding conditions.
- The MVCC schemes have higher overhead but are more resilient, retaining good throughput even in the presence of hotspots and long read-only transactions.
- The optimistic MVCC scheme consistently achieves higher throughput than the pessimistic scheme.

## 8. ACKNOWLEDGMENTS
We thank Phil Bernstein, Marcel van der Holst and Dimitris Tsirogiannis for their contributions. This work was supported in part by a grant from the Microsoft Jim Gray Systems Lab.